\documentclass[aps,prd,showpacs,nofootinbib]{revtex4}
\usepackage{graphicx,color}
\usepackage[latin1]{inputenc}
\usepackage[english]{babel}

\newcommand{\bm}[1]{\mbox{\boldmath $#1$}}
\newcommand{\be}{\begin{equation}}
\newcommand{\ee}{\end{equation}}
\newcommand{\bea}{\begin{eqnarray}}
\newcommand{\eea}{\end{eqnarray}}

\newcommand{\bfk}{\mbox{\boldmath $k$}}

\newcommand{\cupar}{c^\uparrow}

\newcommand{\bfp}{\mbox{\boldmath $p$}}

\newcommand{\la}{\lambda}
\newcommand{\hf}{\hat f}

\def\lsim{\mathrel{\rlap{\lower4pt\hbox{\hskip1pt$\sim$}}\raise1pt\hbox{$<$}}}
\def\gsim{\mathrel{\rlap{\lower4pt\hbox{\hskip1pt$\sim$}}\raise1pt\hbox{$>$}}}
\def\nostrocostruttino#1\over#2{\mathrel{\mathop{\kern 0pt \rlap
{\hbox{$#1$}}} \hbox{\kern-.135em $#2$}}}

\newcommand{\vphi}{\varphi}

\newcommand{\IoneG}{\mathcal{T}_{1}^{g}}

\newcommand{\Ione}{\mathcal{T}_{1}}
\newcommand{\Itwo}{\mathcal{T}_{2}}

\newcommand{\NP}[1]{{\it Nucl.\ Phys.}\ {\bf #1}}
\newcommand{\ZP}[1]{{\it Z.\ Phys.}\ {\bf #1}}
\newcommand{\PL}[1]{{\it Phys.\ Lett.}\ {\bf #1}}
\newcommand{\PR}[1]{{\it Phys.\ Rev.}\ {\bf #1}}
\newcommand{\PRL}[1]{{\it Phys.\ Rev.\ Lett.}\ {\bf #1}}

\newcommand{\EPJ}[1]{{\it Eur.\ Phys.\ J.}\ {\bf #1}}

\begin{document}

\title{Intrinsic parton motion soft mechanisms and the \\
longitudinal spin asymmetry
$\bm{A_{LL}}$ in high energy $\bm{pp\rightarrow \pi X}$}

\author{M.~Anselmino$^1$, M.~Boglione$^1$, U.~D'Alesio$^{2,3}$,
E.~Leader$^{2,4}$, S.~Melis$^{1,2}$, F.~Murgia$^3$}

\vspace{0.5cm}

\affiliation{$^1$ Dipartimento di Fisica Teorica, Universit\`a di Torino and
          INFN, Sezione di Torino, Via P. Giuria 1, I-10125 Torino, Italy\\
$^2$ Dipartimento di Fisica,
Universit\`a di Cagliari, Cittadella Universitaria di Monserrato,
I-09042 Monserrato (CA), Italy\\
$^3$ INFN, Sezione di Cagliari, C.P. 170, I-09042 Monserrato (CA), Italy\\
$^4$ Imperial College London, Prince Consort Road, London SW7 2BW,
U.K.}

\date{\today}

\begin{abstract}

\noindent

The longitudinal double spin asymmetry $A_{LL}$ in the reaction
$pp\rightarrow \pi X$ has been measured at RHIC with extremely interesting
consequences. If the gluon polarization in a proton were as big as needed to
resolve the famous ``spin crisis" then $A_{LL}$ would be large and positive.
Latest RHIC results indicate that $A_{LL}$ is small and disfavour large positive values of the gluon polarization.
We examine whether the soft mechanisms
(Collins, Sivers, Boer-Mulders), essential for generating transverse single spin
asymmetries, have any significant influence on $A_{LL}$, and whether they
could alter the conclusion that the gluon polarization
is necessarily small. It turns out that the contribution from these
effects is essentially negligible.

\end{abstract}


\maketitle

\section{\label{a}Introduction}

Large transverse single spin asymmetries (up to 40\%) have been observed in a
multitude of reactions for over three decades, whereas such asymmetries are tiny
($\lsim 1$\%) in the standard leading twist QCD parton model.
To explain the size of these asymmetries Sivers and Collins~\cite{Sivers:1989cc,Collins:1992kk}
introduced new
soft mechanisms, utilizing, as an essential ingredient, the intrinsic
transverse momentum of partons~\cite{am}.
Later other similar mechanisms were shown to be possible, for example the
Boer-Mulders mechanism~\cite{amst,dan}.

Although these mechanisms were invented in order to
produce transverse asymmetries, it turns out that they also contribute to the
longitudinal double spin asymmetries and to the total cross-section~\cite{X}.
For the latter, it has been shown
that the effect of the soft functions is negligible. However, it was found that
intrinsic transverse momentum {\it per se} significantly affects the value of the
cross-section~\cite{fu}.
That this should be the case at lower energies had already been noticed by Field
and Feynman~\cite{ff}, and later by Vogelsang and Weber~\cite{werner-3} on the
grounds that taking into account intrinsic $k_\perp$
is a particular method of including higher twist corrections.

For the longitudinal asymmetries, the question is much more delicate, for the
following reason.
One of the most important reactions measured at RHIC is the double spin
longitudinal asymmetry $A_{LL}$, which has been found to be very small, and
which has been used, based on a leading twist collinear treatment,
to confirm the growing belief that the gluon polarization is far too small to explain the ``spin crisis in the parton model''~\cite{LA1988}.

Given that  $A_{LL}$ is so
small, and that the implications of this are so important, we felt it
necessary to check whether the soft mechanisms can have a significant
impact, in particular whether they could influence the above conclusion
about the gluon polarization.
We have also checked for any significant
sensitivity in $A_{LL}$ to intrinsic transverse momentum.

The plan of the paper is the following. In Section II we briefly recall the
formalism used for our calculation, which includes the full non-collinear
kinematics of the scattering process. In Section III we present the
``kernels'' for the calculation of each partonic contribution to the polarized
cross-sections. In Section IV we show and discuss our phenomenological
results for the longitudinal double spin asymmetry in inclusive neutral pion
production at RHIC. Finally, in Section V we draw
our conclusions.

\section{Formalism}

Here we simply sketch the main aspects of the formalism; for details of the
approach we refer to~\cite{X}. The longitudinal double spin asymmetry $A_{LL}$
for the reaction $pp\rightarrow \pi X$ is defined as
\be
A_{LL}=\frac{d\sigma^{++}-d\sigma^{+-}}{d\sigma^{++}+d\sigma^{+-}}=
\frac{d\sigma^{++}-d\sigma^{+-}}{2d\sigma^{unp}}\,,
\label{all-def}
\ee
where the labels refer to the helicities of the protons.

The general expression for the differential cross-sections for the polarized
hadronic process $(A,S_A) + (B,S_B) \to C + X $ is given by
\bea
 \frac{E_C \, d\sigma^{(A,S_A) + (B,S_B) \to C + X}}
{d^{3} \bfp_C}
 &= & \sum_{a,b,c,d, \{\la\}}
 \! \! \int \frac{dx_a \,
dx_b \, dz}{16 \pi^2 x_a x_b z^2  s} \;
d^2 \bfk_{\perp a} \, d^2 \bfk_{\perp b}\, d^3 \bfk_{\perp C}\,
\delta(\bm{k}_{\perp C} \cdot \hat{\bm{p}}_c) \, \nonumber \\
 \!\!& \times & \! \! J(\bm{k}_{\perp C})
 \rho_{\la^{\,}_a,
\la^{\prime}_a}^{a/A,S_A} \, \hat f_{a/A,S_A}(x_a,\bfk_{\perp a})
\> \rho_{\la^{\,}_b, \la^{\prime}_b}^{b/B,S_B} \,
\hat f_{b/B,S_B}(x_b,\bfk_{\perp b}) \label{gen1} \nonumber \\
\! \! &\times& \! \! \hat M_{\la^{\,}_c, \la^{\,}_d; \la^{\,}_a, \la^{\,}_b} \,
\hat M^*_{\la^{\prime}_c, \la^{\,}_d; \la^{\prime}_a,
\la^{\prime}_b} \> \delta(\hat s + \hat t + \hat u) \> \hat
D^{\la^{\,}_C,\la^{\,}_C}_{\la^{\,}_c,\la^{\prime}_c}(z,\bfk_{\perp
C})\,,
\eea
which involves a (factorized) convolution of all possible
hard elementary QCD processes, $ab \to cd$, with soft partonic polarized
distribution and fragmentation functions. In Eq.~(\ref{gen1}) $\hat s, \hat t$
and $\hat u$ are the Mandelstam variables for the partonic reactions.
The detailed connection between the hadronic and the partonic kinematical
variables is given in full in Appendix A of Ref.~\cite{X}. A discussion of some
technical details, like, {\it e.g.}, the infrared regulators related to small
partonic scattering angles, can be found, for example, in Ref.~\cite{fu}.

Let us simply recall here, for a better understanding, the physical meaning of
the different factors in Eq.~(\ref{gen1}):

\begin{itemize}
\item
$\rho_{\la^{\,}_a, \la^{\prime}_a}^{a/A,S_A}$ is the helicity density matrix
of parton $a$ inside the polarized hadron $A$, with spin state $S_A$; it
describes the parton polarization. $\hat f_{a/A,S_A}(x_a,\bfk_{\perp a})$ is
the number density (or distribution) of unpolarized partons $a$ inside the
polarized hadron $A, S_A$: each parton carries a light-cone
momentum fraction $x_a$ and a transverse momentum $\bfk_{\perp a}$.
Similarly for parton $b$ inside hadron $B$ with spin $S_B$.
\item
The polarized cross-sections for the elementary
partonic process $(a, s_a) + (b,s_b) \to (c,s_c) + d$ are expressed in terms
of  products of the helicity amplitudes
$\hat M_{\la^{\,}_c, \la^{\,}_d; \la^{\,}_a, \la^{\,}_b}$.
\item
The factor $\hat D^{\la^{\,}_C,\la^{\,}_C}_{\la^{\,}_c,\la^{\prime}_c}(z,
\bfk_{\perp C})$ describes, again in the helicity basis, the fragmentation
process $c \to C + X$, according to which a polarized parton
$c$ fragments into an unpolarized hadron $C$ carrying a light-cone momentum
fraction $z$ and a transverse momentum $\bfk_{\perp C}$.
\item
$J(\bm{k}_{\perp C})$ is a kinematical factor, numerically very close to 1 for
RHIC kinematics.
All details can be found in Ref.~\cite{X}.
Throughout the paper, we work in the $AB$ c.m.~frame, assuming that hadron
$A$ moves along the positive $Z_{cm}$-axis and hadron $C$ is produced in the
$(XZ)_{cm}$ plane, with $(p_C)_{X_{cm}}>0$.
\end{itemize}

Eq.~(\ref{gen1}) is written in a factorized form, separating the soft, long
distance from the hard, short distance contributions. The hard part
is computable in perturbative QCD, while information on the soft one
has to be extracted from other experiments or modeled. As already mentioned
and discussed in Refs.~\cite{fu,X}, such a factorization with non-collinear
kinematics has never been formally proven. Indeed, studies of factorization~\cite{piet,metz,ji,col2}, comparing semi-inclusive deep inelastic scattering
(SIDIS) and Drell-Yan reactions have indicated unexpected modifications of
simple factorization, and the situation for single inclusive particle
production in hadron--hadron collisions is not yet resolved. Thus, our
approach can only be considered as the natural extension of the collinear
case and a reasonable phenomenological model. Of course, the perturbative
calculation of the hard part is only reliable if the hard scale -- in this
case the square of the transverse momentum of the final hadron, $p_T^2$ --
is large enough. It turns out that the data on unpolarized cross-sections in
hadronic collisions at low-intermediate energy scales suggest~\cite{fu} an
average value of $k_\perp^2 \equiv |\bfk _\perp|^2 \simeq 0.64$ (GeV/$c$)$^2$
for the intrinsic transverse momentum of the parton distributions. On the
other hand, both unpolarized hadronic cross-sections at RHIC energies~\cite{Boglione:2007dm}
and the Cahn effect in SIDIS~\cite{sidis} are rather well
reproduced by using $\langle k_\perp^2 \rangle \simeq 0.25$ (GeV/$c$)$^2$.
We shall therefore study how the contributions to $A_{LL}$ depend on the
value of $\langle k_\perp^2 \rangle$.

\section{\label{sigma} Kernels}

As we can see from Eq.~(\ref{gen1}), the computation of the cross-section
corresponding to any polarized hadronic process $(A,S_A) + (B,S_B) \to C + X$
requires the evaluation and integration, for each elementary process
$a+b\to c+d$, of the general kernel
\bea
\Sigma(S_A,S_B)^{ab\to cd}& & \! \! \!= \sum _{\{\lambda\}} \rho_{\la^{\,}_a,
\la^{\prime}_a}^{a/A,S_A} \, \hat f_{a/A,S_A}(x_a,\bfk_{\perp a})
  \rho_{\la^{\,}_b,
\la^{\prime}_b}^{b/B,S_B} \, \hat f_{b/B,S_B}(x_b,\bfk_{\perp b})
\label{kern}  \nonumber \\
&& \times \hat M_{\la^{\,}_c, \la^{\,}_d; \la^{\,}_a, \la^{\,}_b}
\hat M^*_{\la^{\prime}_c,
\la^{\,}_d; \la^{\prime}_a, \la^{\prime}_b} \> \hat
D^{\la^{\,}_C,\la^{\,}_C}_{\la^{\,}_c,\la^{\prime}_c}(z,\bfk_{\perp C})\,.
\eea

While the hadronic process $(A,S_A) + (B,S_B) \to C + X$ takes place,
according to our choice, in the $(XZ)_{cm}$ plane, all the elementary
processes involved, $A(B) \to a(b) + X$, $ab \to cd$ and $c \to C + X$ do not,
since all parton and hadron momenta, $\bfp_a, \, \bfp_b, \, \bfp_C$  have
transverse components $\bfk_{\perp a}, \, \bfk_{\perp b}, \, \bfk_{\perp C}$.
This ``out of $(XZ)_{cm}$ plane'' geometry induces phases in the fragmentation
process, in the distribution functions and in the elementary interactions,
which have to be taken into account. Thus, the independent helicity
amplitudes for the elementary pQCD processes $ab \to cd$, with massless
partons, can be written as~\cite{X}
\be
\hat M_{+,+;+,+}  \equiv \hat M_1^0 \,e^{i\varphi_1} \qquad
\hat M_{-,+;-,+}  \equiv  \hat M_2^0 \,e^{i\varphi_2} \label{phases} \qquad
\hat M_{-,+;+,-} \equiv  \hat M_3^0 \,e^{i\varphi_3},
\ee
where the amplitudes $\hat M^0_{1,2,3}$ are the real planar amplitudes defined
in the partonic $ab \to cd$ c.m.~frame,
\be
\hat M_1^0 \equiv \hat M^0_{+,+;+,+} = \hat M^0_{-,-;-,-} \qquad
\hat M_2^0 \equiv \hat M^0_{-,+;-,+} = \hat M^0_{+,-;+,-} \qquad
\hat M_3^0 \equiv \hat M^0_{-,+;+,-} = \hat M^0_{+,-;-,+} , \label{Mqq}
\ee
as required by parity invariance. The phases $\varphi_{1,2,3}$ are complicated
functions of the polar and azimuthal angles of the transverse momenta,
$\bfk_{\perp a}, \bfk_{\perp b}$ and $\bfk_{\perp C}$, and their explicit
expressions can be found in Ref.~\cite{X}.
The relations
\be
\hat M_{-,-;-,-} = \hat M_{+,+;+,+}^* \qquad
\hat M_{+,-;+,-} = \hat M_{-,+;-,+}^* \qquad
\hat M_{+,-;-,+} = \hat M_{-,+;+,-}^* ,
\ee
follow from Eqs.~(\ref{phases}),~(\ref{Mqq}) and from the fact that the phases
$\varphi_i$ change sign by helicity inversion~\cite{X}.
Note that the $+$ and $-$ subscripts refer to $(+1/2)$ and $(-1/2)$ helicities
for quarks, and to $(+1)$ and $(-1)$ helicities for gluons.
There are eight elementary contributions $ab\to cd$ which we have to consider
separately
\bea
&& q_a q_b \to q_c q_d \,, \quad  g_a g_b \to g_c g_d \,, \nonumber \\
&& q g \to q g \,, \quad  g q \to g q \,,\nonumber \\
&& q g \to g q \,, \quad  g q \to q g \,, \\
&& g_a g_b \to q \bar{q}\,, \quad  q \bar{q} \to g_c g_d \,, \nonumber
\eea
where $q$ can in general be either a quark or an antiquark. The subscripts
$a,b,c,d$ for quarks, when necessary, identify the flavour (only in processes
where different flavours can be present); for gluons, these labels identify
the corresponding hadron ($a \to A, \> b \to B, \> c \to C$). By performing
the explicit sums in Eq.~(\ref{kern}), we obtain the kernels for each of the
elementary processes. Note that the new aspect of our calculation is the
appearance of the phases which is a reflection of the non-collinear
kinematics.

The computation of the denominator/numerator of $A_{LL}$ in Eq.~(\ref{all-def}) requires
the evaluation of the kernels $[\Sigma(+,+)\, \pm \,\Sigma(+,-)]$
respectively. The expressions for the sums of kernels, which are relevant for
the unpolarized cross-section, are given in Ref.~\cite{X}. Here we give in detail
the expressions for the differences. They are calculated from the general
kernel given in Eq.~(\ref{kern}). In the following certain terms are
underlined: these are terms which vanish after integration over the angles
of the momenta $\bfk_{\perp a}, \, \bfk_{\perp b}, \, \bfk_{\perp C}$
in Eq.~(\ref{gen1}); we shall further comment on that at the end of this Section.
$\phi_C^H$ is the azimuthal angle
of the hadron $C$ in the parton $c$ helicity frame and its expression in terms
of the angles of $\bfk_{\perp C}$ is given in Appendix A of Ref.~\cite{X}.
Notice that all angular dependences of the kernels are explicitly extracted
and the parton distribution (PDF) and fragmentation (FF) functions only depend on the
magnitudes of the transverse momentum vectors.

\begin{widetext}
\begin{itemize}

\item
\vskip 3pt
\noindent    $q_a q_b \to q_c q_d$ contribution
\bea
&&\hspace*{-1.cm}[\Sigma(+,+)- \Sigma(+,-)]^{q_aq_b\to q_cq_d} =
\nonumber \\
&&\Delta \hat{f}^{a}_{s_z/+}(x_a, k_{\perp a})
\,\Delta\hat{f}^{b}_{s_z/+}(x_b,k_{\perp b})
\,\left[ \, |{\hat M}_1^0|^2 - |{\hat M}_2^0|^2 - |{\hat M}_3^0|^2 \right]
\,\hat{D}_{C/c}(z,k_{\perp C})\nonumber \\
&+&
\Bigg[\,\Delta \hat{f}^{a}_{s_x/+}(x_a, k_{\perp a})
\,\Delta \hat{f}^{b}_{s_x/+}(x_b, k_{\perp b})\,\cos(\vphi_3-\vphi_2)
\nonumber \\
&&\>\quad
+\underline{\Delta \hat{f}^{a}_{s_y/A}(x_a, k_{\perp a})
\,\Delta \hat{f}^{b}_{s_x/+}(x_b, k_{\perp b})\,\sin(\vphi_3-\vphi_2)}
\,\Bigg]\; (2\,{\hat M}_2^0 \, {\hat M}_3^0)\;\hat{D}_{C/c}(z,k_{\perp C})
\nonumber \\ \label{all-qqqq}
&-& \underline{\hf_{a/A}(x_a, k_{\perp a})
\,\Delta\hat{f}^{b}_{s_x/+}(x_b,k_{\perp b})
\,{\hat M}_1^0 \, {\hat M}_3^0\,\sin(\varphi_1 -\varphi_3 +\phi_C^H)
\,\Delta ^N {\hat D} _{C/\cupar} (z, k_{\perp C})}\,.
\eea
Notice that we have used the relations
$\Delta {\hat f}^a_{s_y/+}(x_a, k_{\perp a})=
 \Delta {\hat f}^a_{s_y/A}(x_a, k_{\perp a})$ and
$\hf_{a/+}(x_a, k_{\perp a})=\hf_{a/A}(x_a, k_{\perp a})$,
see Appendix B of Ref.~\cite{X}.
The channels $q\bar{q}\to q\bar{q}$ etc.~are formally identical
to $qq\to qq$ with amplitudes defined properly in Ref.~\cite{X}.

\item
\vskip 3pt
\noindent  $g_a g_b \to g_c g_d$ contribution
\bea
&&\hspace*{-1.cm}[\Sigma(+,+)- \Sigma(+,-)]^{g_a g_b\to g_c g_d} =
\nonumber \\
&&\Delta \hf^{a}_{s_z/+}(x_a, k_{\perp a})\,
\Delta \hf^{b}_{s_z/+}(x_b, k_{\perp b})
\, \left[ \, |{\hat M}_1^0|^2 - |{\hat M}_2^0|^2 - |{\hat M}_3^0|^2 \right]
\,\hat D _{C/g} (z, k_{\perp C}) \nonumber \\
&+&
\Bigg[\Delta \hf_{\Itwo/+}^{a}(x_a, k_{\perp a})
\,\Delta \hf_{\Itwo/+}^{b}(x_b, k_{\perp b})\,\cos(\vphi_3-\vphi_2)
\nonumber \\
&&+\,
\underline{\Delta\hf_{\Ione/A}^{a}(x_a,k_{\perp a})
\Delta \hf_{\Itwo/+}^{b}(x_b, k_{\perp b})\,\sin(\vphi_3-\vphi_2)}\,
\Bigg]\; (2\,{\hat M}_2^0 \, {\hat M}_3^0)\;\hat{D}_{C/g}(z,k_{\perp C})
\quad\label{all-gggg}  \nonumber \\
&+&\,\underline{\hf_{a/A}(x_a,k_{\perp a})
\,\Delta\hat{f}^{b}_{\Itwo/+}(x_b,k_{\perp b})
\,{\hat M}_1^0 \, {\hat M}_3^0\,\sin(\varphi_1 -\varphi_3 +2\phi_C^H)
\,\Delta ^N {\hat D} _{C/\IoneG} (z, k_{\perp C})}
\eea

\item
\vskip 3pt
\noindent   $q \bar{q} \to g g$ contribution
\bea
&&\hspace*{-1.cm}[\Sigma(+,+)- \Sigma(+,-)]^{q \bar{q} \to g g} =
\nonumber \\
&-&\Delta \hat{f}^{a}_{s_z/+}(x_a, k_{\perp a})
\,\Delta\hat{f}^{b}_{s_z/+}(x_b,k_{\perp b})
\,\left[ \, |{\hat M}_2^0|^2 + |{\hat M}_3^0|^2 \right]
\,\hat{D}_{C/g}(z,k_{\perp C})\nonumber\\
&+&
\Bigg[\,\Delta \hat{f}^{a}_{s_x/+}(x_a, k_{\perp a})
\,\Delta \hat{f}^{b}_{s_x/+}(x_b, k_{\perp b})\,\cos(\vphi_3-\vphi_2)
\nonumber\\
&&\>
+\,\underline{\Delta \hat{f}^{a}_{s_y/A}(x_a, k_{\perp a})
\,\Delta \hat{f}^{b}_{s_x/+}(x_b, k_{\perp b})\,\sin(\vphi_3-\vphi_2)}
\,\Bigg]\; (2 \,{\hat M}_2^0 \, {\hat M}_3^0)\;\hat{D}_{C/g}(z,k_{\perp C})
\label{all-qqgg}
\eea

\item
\vskip 3pt
\noindent  $g_a g_b \to q\bar{q}/\bar{q}q$ contribution
\bea
&&\hspace*{-1.cm}[\Sigma(+,+)- \Sigma(+,-)]^{g_a g_b \to q\bar{q}/\bar{q}q} =
\nonumber \\
&-&\Delta \hf^{a}_{s_z/+}(x_a, k_{\perp a})\,
\Delta \hf^{b}_{s_z/+}(x_b, k_{\perp b})\,
\left[ \, |{\hat M}_2^0|^2 + |{\hat M}_3^0|^2 \right]
\,\hat D _{C/c} (z, k_{\perp C}) \nonumber\\
&+&
\Bigg[\Delta \hf_{\Itwo/+}^{a}(x_a, k_{\perp a})
\,\Delta \hf_{\Itwo/+}^{b}(x_b, k_{\perp b})\,\cos(\vphi_3-\vphi_2)
\nonumber\\
&&+\,
\underline{\Delta\hf_{\Ione/A}^{a}(x_a,k_{\perp a})
\Delta \hf_{\Itwo/+}^{b}(x_b, k_{\perp b})\,\sin(\vphi_3-\vphi_2)}
\,\Bigg]\; (2 \,{\hat M}_2^0 \, {\hat M}_3^0)\;\hat{D}_{C/c}(z,k_{\perp C})
\label{all-ggqq}
\eea

\item
\vskip 3pt
\noindent   $q g \to q g$ contribution
\bea
&&\hspace*{-1.cm}[\Sigma(+,+)- \Sigma(+,-)]^{q g \to q g} =
\nonumber \\
&&\Delta \hat{f}^{a}_{s_z/+}(x_a, k_{\perp a})
\,\Delta\hat{f}^{b}_{s_z/+}(x_b,k_{\perp b})
\,\left[ \, |{\hat M}_1^0|^2 - |{\hat M}_2^0|^2\right]
\,\hat{D}_{C/c}(z,k_{\perp C})
\label{all-qgqg}
\eea

\item
\vskip 3pt
\noindent     $g q \to q g$ contribution
\bea
&&\hspace*{-1.cm}[\Sigma(+,+)- \Sigma(+,-)]^{g q \to q g} =
\nonumber \\
&&\Delta \hat{f}^{a}_{s_z/+}(x_a, k_{\perp a})
\,\Delta\hat{f}^{b}_{s_z/+}(x_b,k_{\perp b})
\,\left[ \, |{\hat M}_1^0|^2 - |{\hat M}_3^0|^2 \right]\,
\hat{D}_{C/c}(z,k_{\perp C}) \nonumber \\
&-&\underline{\hf_{a/A}(x_a, k_{\perp a})
\,\Delta\hat{f}^{b}_{s_x/+}(x_b,k_{\perp b})
\,{\hat M}_1^0 \, {\hat M}_3^0\,\sin(\varphi_1 -\varphi_3 +\phi_C^H)
\,\Delta ^N {\hat D} _{C/\cupar} (z, k_{\perp C})}
\eea

\item
\vskip 3pt
\noindent  $q g \to g q$ contribution
\bea
&&\hspace*{-1.cm}[\Sigma(+,+)- \Sigma(+,-)]^{q g \to g q} =
\nonumber \\
&&\Delta \hf^{a}_{s_z/+}(x_a, k_{\perp a})
\,\Delta \hf^{b}_{s_z/+}(x_b, k_{\perp b})\, \left[ \,
|{\hat M}_1^0|^2 - |{\hat M}_3^0|^2 \right]
\,\hat D _{C/g} (z, k_{\perp C})\nonumber \\
&+&\,\underline{\hf_{a/A}(x_a,k_{\perp a})
\,\Delta\hat{f}^{b}_{\Itwo/+}(x_b,k_{\perp b})
\,{\hat M}_1^0 \, {\hat M}_3^0\,\sin(\varphi_1 -\varphi_3 +2\phi_C^H)
\,\Delta ^N {\hat D} _{C/\IoneG} (z, k_{\perp C})}
\eea

\item

\vskip 3pt
\noindent $g q \to g q$ contribution
\bea
&&\hspace*{-1.cm}[\Sigma(+,+)- \Sigma(+,-)]^{g q \to g q} =
\nonumber \\
&&\Delta \hf^{a}_{s_z/+}(x_a, k_{\perp a})\,
\Delta \hf^{b}_{s_z/+}(x_b, k_{\perp b})
\,\left[ \, |{\hat M}_1^0|^2 - |{\hat M}_2^0|^2 \right]
\,\hat D _{C/g} (z, k_{\perp C}) \;.
\label{all-gqgq}
\eea

\end{itemize}
\end{widetext}

The physical content of the above expressions is interesting.
First note the complete formal symmetry between the $qq\to qq$ kernel in
Eq.~(\ref{all-qqqq}) and the $gg\to gg$ kernel in
Eq.~(\ref{all-gggg}).
These kernels contain the largest variety of contributions, and the
kernels for all the other partonic processes can be formally read off from
these by the suppression of certain terms.

In the second line of both expressions, Eq.~(\ref{all-qqqq}) and
Eq.~(\ref{all-gggg}), we recognize the product of the $k_{\perp}$-dependent
helicity distributions, $\Delta \hat{f}^{q}_{s_z/+}(x_q, k_{\perp q}) \equiv
\Delta q(x_q, k_{\perp q})$ and $\Delta \hat{f}^{g}_{s_z/+}(x_g, k_{\perp g})
\equiv \Delta g(x_g, k_{\perp g})$ for quarks and gluons respectively,
and the unpolarized fragmentation function $\hat{D}_{C/c}(z,k_{\perp C})$,
with no azimuthal phases.
In the third line of Eq.~(\ref{all-qqqq}) we have two parton distribution
functions, $\Delta \hat{f}^q_{s_x/+}(x, k_{\perp})$, referring to quarks
transversely polarized, along the $x$-axis, inside longitudinally polarized
nucleons, coupled to the unpolarized fragmentation function.
Analogously, in the third line of Eq.~(\ref{all-gggg})
we have two parton distribution
functions, $\Delta \hf^{g}_{\Itwo/+}(x, k_{\perp})$, which are related to the
linear polarization of a gluon inside a longitudinally polarized nucleon.
Correspondingly the fourth line of Eq.~(\ref{all-qqqq}) refers to one quark
transversely polarized along the $x$-axis inside a longitudinally polarized
nucleon and the other, $\Delta \hat{f}^{q}_{s_y/A}(x, k_{\perp})$,
transversely polarized along the $y$-axis inside an unpolarized  nucleon --
the latter is the Boer-Mulders function -- coupled to the unpolarized
fragmentation function.
Analogously, in the fourth line of Eq.~(\ref{all-gggg})
we have $\Delta \hf^g_{\Itwo/+}(x, k_{\perp})$
and the ``Boer-Mulders-like'' gluon function,
$\Delta\hf_{\Ione/A}^{g}(x,k_{\perp})$, referring to a linearly polarized
gluon inside an \textit{unpolarized} nucleon. For a more complete explanation
of the physical meaning of these functions see Appendix B of Ref.~\cite{X}.
Finally, the last line of Eq.~(\ref{all-qqqq}) contains the Collins
fragmentation function, $\Delta ^N {\hat D} _{C/\cupar} (z, k_{\perp C})$,
coupled to an unpolarized parton density and a transversely polarized one.
In the case of the gluon, in the last line of Eq.~(\ref{all-gggg}), there
appears a gluonic analogue of the Collins fragmentation function,
$\Delta ^N {\hat D} _{C/\IoneG} (z, k_{\perp C})$, describing the
fragmentation of a linearly polarized gluon into an unpolarized hadron.

Ignoring the underlined terms which vanish upon integration, we see that
compared to the standard collinear approach, we have  extra contributions
involving quarks polarized transversely along their $x$-axis in a
longitudinally polarized nucleon, appearing in Eqs.~(\ref{all-qqqq}),
(\ref{all-qqgg}) and contributions involving linearly polarized gluons
inside a longitudinally polarized nucleon, appearing in Eqs.~(\ref{all-gggg}),
(\ref{all-ggqq}). Notice that the processes in
Eqs.~(\ref{all-qgqg})-(\ref{all-gqgq}), initiated by quark-gluon elementary
scattering, get contributions only from the usual terms, which survive in the
collinear case.

The demonstration of the vanishing upon angular integration
of the underlined terms in Eqs.~(\ref{all-qqqq})-(\ref{all-gqgq})
requires a detailed study of the kinematics and of the
relationships between the angular integration variables
appearing in $\bfk_{\perp a}, \, \bfk_{\perp b}, \, \bfk_{\perp C}$
in Eq.~(\ref{gen1}) and the phase variables $\varphi_{1,2,3}$
and $\phi_C^H$~\cite{X}. We have also numerically checked that
this is indeed the case.

Notice that a parity transformation implies $\varphi_i \to -\varphi_i$ $(i = 1,2,3)$
and $\phi_C^H \to -\phi_C^H$. Thus the odd $\sin$ terms in
Eqs.~(\ref{all-qqqq})-(\ref{all-gqgq}) must vanish if parity is conserved.

Another simple, but interesting example of such a vanishing can be obtained by
considering, within the same formalism, the expression of the kernels for the
longitudinal \textit{single spin} asymmetry $A_L$, which we know \textit{must}
vanish in a parity conserving theory. The kernels themselves are \textit{not}
zero, but under integration do vanish. This is another very stringent test of
the correctness of our formalism. For $A_L$, for the partonic channel
$q_a q_b \to q_c q_d$, we have for the numerator of the longitudinal
single spin asymmetry the following expression:
\bea
&&\hspace*{-1.cm}[\Sigma(+,0)- \Sigma(-,0)]^{q_aq_b\to q_cq_d}\label{eq:AL} =
\nonumber \\
&& \underline{\Delta \hat{f}^{a}_{s_x/+}(x_a, k_{\perp a})
\,\Delta \hf^b_{s_y/B}(x_b,k_{\perp b})
\,(2{\hat M}_2^0 \, {\hat M}_3^0)\,\sin(\vphi_3-\vphi_2)
\,\hat{D}_{C/c}(z,k_{\perp C})}\nonumber\\
&&\underline{-\Delta \hat{f}^{a}_{s_x/+}(x_a, k_{\perp a})
\,\hf_{b/B}(x_b, k_{\perp b})\,\sin(\vphi_1 - \vphi_2 + \phi_C^{H})
\,{\hat M}_1^0 \, {\hat M}_2^0\,
\Delta ^N {\hat D} _{C/\cupar} (z, k_{\perp C})}\,,
\eea
and, again, all terms -- being odd functions of $\varphi_i$ and
$\phi_C^H$ -- vanish, as they should, upon angular integration.

\newpage

\section{Phenomenology: $\bm{A_{LL}}$ at RHIC}

The longitudinal double spin asymmetry $A_{LL}$ for inclusive neutral pion and jet production in proton-proton scattering at $\sqrt{s}=200$ GeV has been
measured at RHIC in various runs, respectively by the PHENIX~\cite{phenix1,Adler:2006bd,Adare:2007dg} and STAR~\cite{star1,Abelev:2007vt} Collaborations.
The first published PHENIX experimental data~\cite{phenix1} showed results
for $A_{LL}$ at mid rapidity compatible with negative values.
This was quite puzzling, since $A_{LL}$ is a positive quantity  in the
collinear parton model~\cite{Jager:2003ch},
at least at low $p_T$ where it is dominated by $gg\to gg$ elementary
scattering processes, see Eq.~(\ref{all-gggg}) in which it can be shown that
$|{\hat M}_1^0|^2 - |{\hat M}_2^0|^2 - |{\hat M}_3^0|^2>0$.
More recent and precise data from both
collaborations~\cite{Adler:2006bd,star1} exclude the possibility of a large and
negative $A_{LL}$: in two subsequent RHIC runs, Run $5$~\cite{Abelev:2007vt,Adare:2007dg} and Run $6$
(results from Run 6 have only been presented as ``preliminary''~\cite{aoki,saito}),
they confirm and reinforce
the statement that $A_{LL}$ is very small and compatible with zero over the whole $p_T$ range covered.

An earlier comparison of present RHIC data with collinear next-to-leading order (NLO) QCD calculations of $A_{LL}$~\cite{Jager:2002xm} disfavoured large positive values for $\Delta g$, definitely excluding scenarios where
$\Delta g$ is as large as the unpolarized gluon distribution function, $g$, at low scale.
Instead the data were in better agreement with the predictions
obtained by assuming $\Delta g = 0$ or even $\Delta g =-g$ at the initial
scale~\cite{stratmann-vogelsang}.
A recent statistical analysis shows that
the PHENIX Run 5 data are compatible with both $\Delta g = 0$ and the
``standard'' GRSV parametrization~\cite{GRSV2000},
while it rules out the $\Delta g = -g$ (at the initial scale) hypothesis~\cite{Adare:2007dg}.
A newest update of this analysis,
which includes the preliminary data from PHENIX Run 6, favours the
$\Delta g = 0$ scenario over the standard GRSV~\cite{saito}.
(Note that in this Section we have
adopted the common,
short-hand notation $\Delta {f}^q_{s_z/+} \equiv \Delta q$ and
$\Delta {f}^g_{s_z/+} \equiv \Delta g$
for the helicity distribution functions,
while ${f}_{q/p}\equiv q$ and ${f}_{g/p}\equiv g$ for the
unpolarized distribution functions, for quarks and gluons respectively).

Our goal is to explore whether the new mechanisms permitted by the presence of
partonic intrinsic transverse momenta, obtained in a general and fully
non-collinear kinematics, could affect the above conclusions, which are based
on the analysis of $A_{LL}$ in the collinear configuration, i.e. taking into
account only the terms proportional to $\Delta q(x)$ and $\Delta g(x)$.
Could the ``new'' contributions shown in Eqs.~(\ref{all-qqqq})-(\ref{all-ggqq})
turn the longitudinal double spin asymmetry $A_{LL}$ into a very small (or even
slightly negative) quantity without the need to assume $\Delta g$ to be zero
or negative?

We have studied $A_{LL}$ at RHIC, for the PHENIX kinematics,
$\sqrt{s}=200$ GeV and $|\eta|<0.35$ (numerical calculations are performed at
$\eta=0$) and evaluated each separate contribution to $A_{LL}$, according to
Eqs.~(\ref{all-qqqq})-(\ref{all-gqgq}).
Since we have no knowledge of the parton densities
$\Delta \hat{f}^{q}_{s_x/+}$ and $\Delta \hf_{\Itwo/+}^{g}$
we maximized them in order to see whether, in principle, they can have a
significant effect on $A_{LL}$.
We thus used for them the corresponding unpolarized parton densities and adjusted
the signs so that all contributions add up coherently.

For the helicity distributions we have used the sets GRSV2000~\cite{GRSV2000}
and LSS05~\cite{LSS05}.
The unpolarized cross-section and the maximized
contributions to the numerator of $A_{LL}$ have been calculated
using the GRV98 set~\cite{GRV98} and the MRST01 set~\cite{MRST01}
respectively.
For the fragmentation functions we have used the KKP set~\cite{KKP}
and, for comparison, the Kretzer set~\cite{Kretzer:2000yf}.
The transverse momentum dependence has been
included by means of a factorized Gaussian smearing, for all the parton
distribution  and fragmentation functions
\bea
\hat{f}(x,k_\perp)&=&
f(x)\;\frac{e^{-k^2_\perp/\langle k^2_\perp\rangle}}
{\pi\langle k^2_\perp\rangle}\,,\\
\hat{D}(z,k_{\perp C})&=&D(z)\;\frac{e^{-k^2_{\perp C}/\langle k^2_{\perp C}\rangle}}
{\pi\langle k^2_{\perp C} \rangle}\,,
\eea
with a constant and flavour independent parameter
$\sqrt{\langle k^2_{\perp} \rangle}\equiv k_{0}$, assumed to be the same for
all quark flavours and for gluons;
we shall study the effect of changes in the value of
$k_{0}$. Guided by our previous work,
we compared the results obtained using three different values for $k_{0}$:
$k_{0}=0.8$ GeV/$c$ from studies on the unpolarized $pp$ scattering
cross-sections and single spin asymmetries~\cite{fu},
$k_{0}=0.5$ GeV/$c$ from fitting the Cahn effect in SIDIS~\cite{sidis},
and $k_0=0.01$ GeV/$c$ to recover the collinear configuration.
For the fragmentation functions, we take
$\langle k^2_{\perp C} \rangle = \langle k^2_\perp\rangle$ everywhere.
We have checked that variations in $\langle k^2_{\perp C} \rangle$
induce negligible changes in $A_{LL}$.

It turns out that the new non-collinear soft contributions containing the PDFs
$\Delta \hat{f}^q_{s_x/+}(x, k_{\perp})$ and
$\Delta \hf^g_{\Itwo/+}(x, k_{\perp})$, even if maximized,
are totally negligible.
In fact, in the RHIC kinematical regime considered their maximized contribution
does not exceed, in the lowest $p_T$ range, few percent of the usual terms
(already present in the collinear case), becoming much smaller at larger $p_T$.
We have checked that this result remains true also at lower energies.
Let us remark that a similar situation holds also for the unpolarized
cross-section~\cite{fu}. Although the two additional terms (with respect
to that already present in the collinear case) are of course different
in this case~\cite{X}, involving respectively the convolution of two
Boer-Mulders functions with an unpolarized fragmentation function and
the convolution of a Boer-Mulders function and an unpolarized distribution
with the Collins fragmentation function, their total maximized contribution
reaches at most 1\% of the usual term, being even smaller on the average.

\begin{figure}
\includegraphics[width=0.38\textwidth,angle=-90]{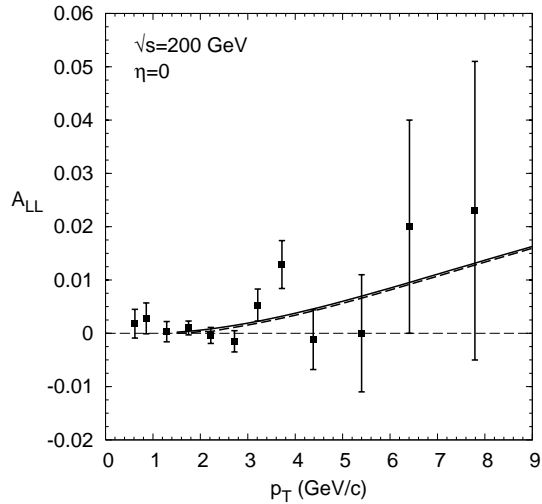}
\caption{$A_{LL}$ for the process $pp\to\pi^0 X$ at $\sqrt{s}=200$ GeV and
$\eta=0$, plotted as a function of $p_T$,
calculated with different choices of
$\sqrt{\langle k^2_{\perp}\rangle}\equiv k_0$
in the PDF/FFs, compared to
PHENIX data, Run 5~\cite{Adare:2007dg}.
The solid line corresponds to the choice
$k_0=0.01$ GeV/$c$ in both PDFs and FFs.
The dashed line corresponds to
$k_0=0.8$ GeV/$c$ in PDF/FFs.
The PDF sets are LSS05~\cite{LSS05} and MRST01~\cite{MRST01},
the FF set is KKP~\cite{KKP} and the factorization
scale is $Q=p_T$. Notice that
the changes in $A_{LL}$ induced by varying the value of
$\langle k^2_{\perp}\rangle$ are
 much smaller than those obtained by choosing different sets of distribution
functions and/or factorization scales, see Fig.~\ref{Tot-01}.
}\label{fig1}
\end{figure}

Therefore, we conclude that there is no way for the extra contributions
induced by the presence of partonic intrinsic transverse momenta to alter the
size of $A_{LL}$. We have checked that this conclusion is not sensitive to the
choice of the mean intrinsic transverse momentum $k_0$.
In fact, Fig.~\ref{fig1} shows that in general $A_{LL}$ depends very little
on the different choices of $k_0$; in particular, $A_{LL}$ decreases
when increasing the width of the gaussian, but compared to data
this variation is quite negligible. This result can be understood because the
$k_\perp$ dependence is given by the same gaussian for all distribution and
fragmentation functions and at mid rapidity the $\hat M$ amplitudes depend
very mildly on  $k_\perp$.

\begin{figure}
\includegraphics[width=0.38\textwidth,angle=-90]{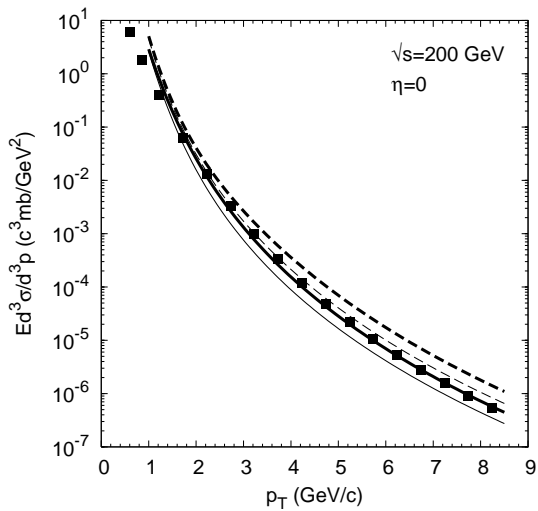}
\caption{The invariant unpolarized cross-section for the process $pp\to\pi^0 X$ at $\sqrt{s}=200$ GeV and
$\eta=0$, plotted as a function of $p_T$, calculated with different FF sets and factorization scales.
The thick, solid and dashed lines correspond to the choice of the KKP FF set~\cite{KKP}, at the factorization scale  $Q=p_T$ and $Q=p_T/2$ respectively. The thin, solid and dashed lines correspond to the choice of the Kretzer FF set~\cite{Kretzer:2000yf}, at the factorization scale  $Q=p_T$ and $Q=p_T/2$ respectively.
The PDF set is GRV98~\cite{GRV98}. $\sqrt{\langle k^2_{\perp}\rangle}\equiv k_0=0.5$ GeV$/c$ for both PDFs and FFs.
The experimental data are from the PHENIX collaboration at RHIC, Run 5~\cite{Adare:2007dg}.}
\label{unp-xsec}
\end{figure}

It is interesting to notice that the corresponding unpolarized cross-section is also almost independent of the value assigned to the average intrinsic transverse momentum $k_0$, while it turns out to be more
sensitive to the choice of the factorization scale and of the fragmentation function set,
as we show in Fig.~\ref{unp-xsec}, consistently with the NLO collinear pQCD calculations.
The comparison with PHENIX data~\cite{Adare:2007dg} is well satisfactory.
The solid lines correspond to the factorization scale $Q=p_T$, the dashed lines to $Q=p_T/2$, using the GRV98~\cite{GRV98} PDF set and the KKP~\cite{KKP} (thick lines) or the Kretzer~\cite{Kretzer:2000yf} (thin lines) FF sets. Results for the STAR and BRAHMS kinematical regimes at $\sqrt{s}=200$ GeV can be found in Ref.~\cite{Boglione:2007dm} and show similar agreement with data when adopting the same average
$k_\perp$'s as in Ref.~\cite{sidis}.

\begin{figure}
\includegraphics[width=0.38\textwidth,angle=-90]{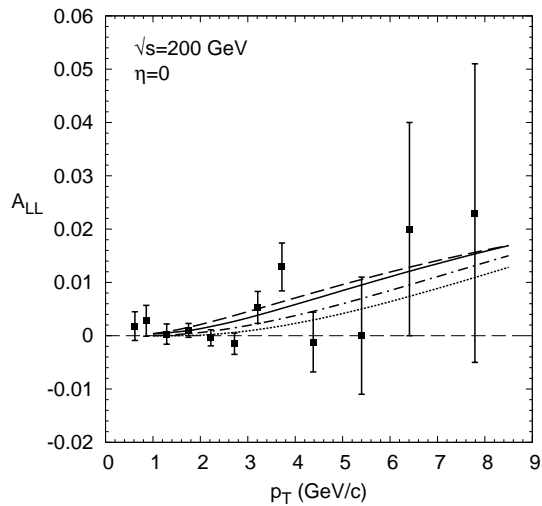}
\caption{%
$A_{LL}$ for the process $pp\to\pi^0 X$ at $\sqrt{s}=200$ GeV and
$\eta=0$, plotted as a function of $p_T$, calculated with different
PDF sets and factorization scales.
The solid and dashed lines correspond to the choice
of PDFs GRSV2000/GRV98, at the factorization scale  $Q=p_T/2$ and $Q=p_T$ respectively.
The dotted and dash-dotted lines correspond the choice
of PDFs LSS05/MRST01, at the factorization scale $Q=p_T/2$ and $Q=p_T$ respectively.
The FF set is KKP.
$\sqrt{\langle k^2_{\perp} \rangle}=0.01$
GeV/$c$ for both PDF/FF. The experimental data are from the PHENIX
collaboration at RHIC, Run 5~\cite{Adare:2007dg}. }\label{Tot-01}
\end{figure}

Contrary to what happens for the $k_\perp$ dependence, $A_{LL}$ is sensitive to the choice of the PDF set and of the scale.
In Fig.~\ref{Tot-01} we show $A_{LL}$ calculated in
an almost collinear configuration, $\sqrt{\langle k^2_{\perp}
\rangle} = 0.01 $ GeV/$c $, and for two choices of scale, $Q=p_T$
and $Q=p_T/2$. Using the LSS05/MRST01
 PDFs, the dotted line corresponds
to $Q=p_T/2$ and the dash-dotted line to $Q=p_T$. Using the
 GRSV2000/GRV98  PDFs,
the solid line corresponds to $Q=p_T/2$, and the
dashed line to $Q=p_T$. As can be seen, the variations induced by different
choices of PDF sets and scale are quite large, larger than those produced by
changes in the $k_0$ value; nevertheless, all these curves are compatible with present
experimental data (we have checked that these same conclusions hold also when adopting $Q=2 p_T$).
However, very precise data on $A_{LL}$ in the future might be able to distinguish between
various sets of PDFs. Data collected at different energies~\cite{aoki}
will also be very useful to cover presently unexplored regions of the Bjorken $x$ variable.

Concerning the dependence of $A_{LL}$ on the set of fragmentation functions, we have checked, adopting again the KKP and Kretzer sets, that this is almost negligible over the whole $p_T$ range considered. Only at the largest $p_T$ values, where $A_{LL}$ data show large experimental errors, there is some residual dependence. This result can be understood, since both the numerator and the denominator of $A_{LL}$ contain the unpolarized FF.
%
%
%
%
%
%
%
%

\section{conclusions}

We have examined, at leading order in perturbative QCD, the effect on the
longitudinal double spin asymmetry $A_{LL}$ of allowing the partons to
have non-zero intrinsic transverse momentum, and of including in
$A_{LL}$ the contributions arising from the new soft functions
that play a crucial role in transverse single spin asymmetries. The study
was carried out in the hope that such effects might negate the
conclusion that the very small measured values of $A_{LL}$
automatically imply that the polarized gluon density is very
small. Our analysis indicates that the contribution from these
effects is negligible and we are forced, at the present stage,
to accept the conclusion that the polarized gluon density is much too small to
explain the ``spin crisis in the parton model"~\cite{LA1988}.

\acknowledgments
We thank W.~Vogelsang and H.~Avakian for useful discussions.
We acknowledge support of the European Community - Research Infrastructure
Activity under the FP6 ``Structuring the European Research Area''
program (HadronPhysics, contract number RII3-CT-2004-506078).
M.A.~and M.B.~acknowledge partial support by MIUR under
Cofinanziamento PRIN 2006.
E.L.~is grateful to Dipartimento di Fisica, Universit\`a di
Cagliari for its generous hospitality.

\end{document}